\documentclass[twocolumn,showpacs,amsmath,amssymb]{revtex4}

\usepackage{graphicx}% Include figure files
\usepackage{dcolumn} % Align table columns on decimal point
\usepackage{bm}      % bold math
\usepackage{latexsym}
\usepackage{color}
\begin{document}

\title{Turbulent magnetic dynamo excitation at low magnetic Prandtl number}

\author{Pablo D. Mininni}
\affiliation{National Center for Atmospheric Research, 
             P.O. Box 3000, Boulder, Colorado 80307}

\date{\today}

\begin{abstract}
Planetary and stellar dynamos likely result from turbulent motions in 
magnetofluids with kinematic viscosities that are small compared to their 
magnetic diffusivities. Laboratory experiments are in progress to produce 
similar dynamos in liquid metals. This work reviews recent computations 
of thresholds in critical magnetic Reynolds number above which dynamo 
amplification can be expected for mechanically-forced turbulence (helical 
and non-helical, short wavelength and long wavelength) as a function of 
the magnetic Prandtl number $P_M$. New results for helical forcing are 
discussed, for which a dynamo is obtained at $P_M=5\times10^{-3}$. The fact 
that the kinetic turbulent spectrum is much broader in wavenumber space 
than the magnetic spectrum leads to numerical difficulties which are 
bridged by a combination of overlapping direct numerical simulations and 
subgrid models of magnetohydrodynamic turbulence. Typically, the critical 
magnetic Reynolds number increases steeply as the magnetic Prandtl number 
decreases, and then reaches an asymptotic plateau at values of at most a 
few hundred. In the turbulent regime and for magnetic Reynolds numbers 
large enough, both small and large scale magnetic fields are excited. 
The interactions between different scales in the flow are also discussed.
\end{abstract}

\maketitle

\section{\label{sec:intro}INTRODUCTION}

Plasmas in stellar interiors and conducting fluids in planetary 
cores are characterized by a magnetic Prandtl number $P_M$ 
(the ratio of the kinematic viscosity $\nu$ to the magnetic 
diffusivity $\eta$) much smaller than one. As a few examples, 
the magnetic Prandtl number in the solar convective region 
is estimated to be $P_M \approx 10^{-5}-10^{-6}$ \cite{Parker79}, 
and in the Earth's core $P_M \approx 10^{-5}$. Liquid sodium 
experiments are also characterized by small values of $P_M$.

While numerical simulations of dynamo action in these 
objects are available, the large values of the kinetic ($R_V$) 
and magnetic ($R_M$) Reynolds numbers forbid a study using 
realistic values of $P_M$. Simulations of the geodynamo 
\cite{Glatzmaier99} or the solar convective region 
\cite{Cattaneo99} are often done for $P_M \sim 1$. While the 
proper separation of the kinetic and magnetic dissipation scales 
cannot be achieved in these simulations, values of $P_M$ much 
smaller than one can be reached under more idealized conditions. 
Pseudospectral methods in periodic boxes give an excellent tool 
to study the behavior of magnetohydrodynamic (MHD) 
turbulence in the regime $P_M<1$. The further assumption of 
incompressibility allows for an extra gain in computer power. 
Paralellized pseudospectral codes have reached for hydrodynamic 
turbulence resolutions of $4096^3$ grid points, and Taylor 
Reynolds numbers of $R_\lambda \approx 1200$ \cite{Kaneda03}. 
These methods are conservative and nondispersive, being well 
suited for the exploration of turbulent flows in regimes hard 
to explore in the laboratory. And under some circumstances, 
the range of values of $P_M$ can be extended using 
subgrid scale (SGS) models.

In this work, we review recent results from simulations of helical 
and non-helical dynamos at $P_M<1$ using pseudospectral codes in 
periodic boxes \cite{Ponty05,Mininni05c,Mininni05e}. To extend the 
range in $P_M$ in the simulations, SGS models were used in these 
works. We discuss some of these models with particular emphasis 
in the Lagrangian Average MHD (LAMHD) equations 
\cite{Holm02a,Holm02b,Montgomery02,Mininni05a,Mininni05b}. 
In addition, and to validate results from SGS models, new results 
from direct numerical simulations (DNS) with resolutions of $1024^3$ 
grid points are presented. 

We focus on the properties of the magnetic field in the kinematic 
dynamo regime, when the intensity of the magnetic field is small 
and the effect of the Lorentz force can be neglected. In particular, 
we discuss the behavior of the threshold in $R_M$ for dynamo action 
as $P_M$ is decreased. That is to say, given a hydrodynamic state 
and an arbitrary small magnetic perturbation, what is the minimum 
value of $R_M$ (or maximum value of $\eta$, in some convenient set 
of dimensionless units) to have a dynamo instability such that the 
system reaches after a finite time a magnetohydrodynamic steady state. 
Below this threshold, the magnetic perturbation is damped and the 
final state of the system is hydrodynamic.

Two flows have recently been studied in this context: the flow 
resulting from Taylor-Green forcing \cite{Ponty05,Mininni05c}, and 
the result of Roberts forcing \cite{Feudel03,Mininni05e}. The first 
case corresponds to a flow with no net helicity that gives large 
scale dynamo action, while the former studies are for a helical 
flow where only small scale dynamo action is permitted by 
introducing mechanical energy in the largest available scale. 
In a different context, for isotropic, homogeneous, 
and delta-correlated in time external forcing, the problem has 
also been studied in Refs. \cite{Haugen04,Schekochihin05}. 
In addition to reviewing this results, we compare them against new 
simulations using Arn'old-Childress-Beltrami (ABC) forcing with energy 
injection at intermediate scales. This is a helical case where large 
scale magnetic amplification is allowed. For this forcing, values of 
$P_M$ down to $5\times10^{-3}$ are reached. The results obtained for 
such a low value of $P_M$ are expected to be of relevance for 
astrophysical and geophysical applications, as well as for laboratory 
dynamos.

For all cases where a large scale flow is present, dynamo action 
is observed to persist at the smallest values of $P_M$ that can 
be reached. Moreover, for values of $P_M$ smaller than $\sim 0.1$ 
an independence of the threshold with $P_M$ 
is observed. While for the Taylor-Green (non-helical) forcing 
and the Roberts forcing (helical, but with magnetic amplification 
only at small scales) a sharp increase in the critical parameter 
is observed before reaching the asymptotic regime, in the ABC 
case almost no such increase is found.

The structure of the paper is as follows. In Sec. 
\ref{sec:equations} we present the equations, and the several 
forcing functions used. We also describe the code and the SGS 
models. Section \ref{sec:thresholds} discusses the thresholds for 
dynamo action for the several flows, and Sec. \ref{sec:scales} 
discusses the role played by the different scales in the problem. 
Finally, Sec. \ref{sec:conclusions} presents the conclusions.

\section{\label{sec:equations}EQUATIONS}

\subsection{The MHD equations}

In dimensionless Alfv\'enic units, the incompressible MHD equations 
are
\begin{eqnarray}
\frac{\partial {\bf v}}{\partial t} + {\bf v \cdot \nabla v} &=& 
    -{\bf \nabla} {\mathcal P} + {\bf j \times B} + \nu \nabla^2 {\bf v} + 
    {\bf f} , \label{eq:momentum} \\
\frac{\partial {\bf B}}{\partial t} + {\bf v \cdot \nabla B} &=&
    {\bf B \cdot \nabla v} + \eta \nabla^2 {\bf B} ,
    \label{eq:induc}
\end{eqnarray}
with ${\bf \nabla \cdot v} = {\bf \nabla \cdot B} = 0$. Here ${\bf v}$ 
is the velocity field, and ${\bf B}$ is the magnetic field, related to 
the electric current density ${\bf j}$ by 
${\bf j} = {\bf \nabla \times B}$. ${\mathcal P}$ is the normalized 
pressure-to-density ratio, obtained by solving the Poisson equation 
that results from taking the divergence of Eq. (\ref{eq:momentum}) 
and using the condition ${\bf \nabla \cdot v} = 0$.

In Eq. (\ref{eq:momentum}) ${\bf f}$ is an external force. Since in 
the incompressible framework we lack the usual energy sources in 
astrophysics and geophysics (e.g. thermal convection), we will set 
${\bf f}$ to generate a large scale flow, and let the instabilities of 
the flow generate turbulent fluctuations. The expressions for 
${\bf f}$ considered are discussed in the following subsection.

Equations (\ref{eq:momentum}) and (\ref{eq:induc}) are solved using a 
standard pseudospectral method \cite{Orszag72,Canutto} in a three 
dimensional periodic box. The code uses projection into the Fourier base 
to compute spatial derivatives, and Runge-Kutta of adjustable order 
to evolve the equations in time. The $2/3$-rule for dealiasing is 
used. As a result, if $N$ grid points are used in each direction, 
the maximum wavenumber solved by the code is $k_{max} = N/3$. To make 
use of parallel computers, the three dimensional Fourier transform 
has to be parallelized efficiently, using a methodology as described 
e.g. in Refs. \cite{Dmitruk01,Gomez05}.

All Reynolds numbers discussed are based on the flow integral 
scale 
\begin{equation}
L = 2\pi \int{E_V(k) k^{-1} \, dk}/E_V
\end{equation}
where $E_V(k)$ is the kinetic energy spectrum, and 
$E_V=\int{E_V(k) dk}$ is the total kinetic energy. Given the r.m.s. 
velocity $U$, the kinetic Reynolds number is $R_V=LU/\nu$ and 
the magnetic Reynolds number is $R_M=LU/\eta$. The magnetic 
Prandtl number is $P_M=R_M/R_V$. We also 
define the large scale eddy turnover time as $T=L/U$. 
The kinetic and magnetic dissipation wavenumbers in the turbulent 
regime are given by $k_\nu=(\epsilon/\nu^3)^{1/4}$ and 
$k_\eta=(\epsilon/\eta^3)^{1/4}$ respectively, where $\epsilon$ is 
the energy injection rate. In all simulations, these wavenumbers 
were smaller than the maximum resolved wavenumber $k_{max}$.

The strategy is to turn on the forcing at $t=0$ and allow the 
code to run for a time as a hydrodynamic code, with the magnetic field 
set to zero. Once a stationary state is reached, the magnetic field is 
seeded with randomly-chosen Fourier coefficients and allowed to 
amplify or decay.

\subsection{External forcing}

Several expressions for the external forcing were studied in the 
context of dynamo action. Refs. \cite{Ponty05,Mininni05c} considered 
Taylor-Green \cite{Taylor37} forcing
{\setlength\arraycolsep{2pt}
\begin{eqnarray}
{\bf f}_{\rm TG} &=& \left[ \sin(k_{\rm TG} x) \cos(k_{\rm TG} y) 
     \cos(k_{\rm TG} z) \hat{x} - \right. {} \nonumber \\
&& {} \left. - \cos(k_{\rm TG} x) \sin(k_{\rm TG} y) 
     \cos(k_{\rm TG} z) \hat{y} \right] ,
\label{eq:TG}
\end{eqnarray}}
\noindent with wavenumber $k_{\rm TG}=2$ (this election gives a peak in 
the kinetic energy at $k\approx3$). If Eq. (\ref{eq:TG}) is used as an initial 
condition for the velocity field ${\bf v}$, it is found that the flow is not a 
solution of the Euler's equation and though highly symmetric, it leads to the 
rapid development of small spatial scales. The forcing is also 
non-helical, in the sense that 
${\bf f}_{\rm TG} \cdot \nabla \times {\bf f}_{\rm TG}$ is zero at 
every point in space. Helical fluctuations can however appear at scales 
smaller than the forcing scale.

Ref. \cite{Feudel03,Mininni05e} considered the Roberts flow 
\cite{Roberts72,Dudley89} as the expression for the external force 
${\bf f}$:
{\setlength\arraycolsep{2pt}
\begin{eqnarray}
{\bf f}_{\rm R} &=& \left[ g \sin(k_{\rm R} x) \cos(k_{\rm R} y) \hat{x} 
     -g \cos(k_{\rm R} x) \sin (k_{\rm R} y) \hat{y} + \right. {} 
     \nonumber \\
&& {} \left. + 2 f \sin(k_{\rm R} x) \sin(k_{\rm R} y) \hat{z} \right] ,
\label{eq:Roberts}
\end{eqnarray}}
\noindent with the choice $k_{\rm R} = 1$. Since in that case mechanical 
energy is injected in the largest available scale, magnetic excitations can 
only grow at scales smaller than the energy injection scale. The 
coefficients $f$ and $g$ are arbitrary and their ratio determines the extent 
to which the flow excited will be helical. We will 
concentrate here upon the case $f = g$. In this case, the 
forcing injects helicity in the flow, although not maximally. If used 
as an initial condition, the velocity field is an exact solution of 
Euler's equation. As a result, the development of small scale 
fluctuations only takes place above a certain threshold 
in the kinetic Reynolds number $R_V$. For values of $R_V$ smaller 
than this threshold, there is an exact laminar solution of the 
hydrodynamic equations given by ${\bf v} = {\bf f}/(\nu k_{\rm R}^2)$.

We compare the results from these two forcing functions 
with the ABC forcing
{\setlength\arraycolsep{2pt}
\begin{eqnarray}
{\bf f}_{\rm ABC} &=& \left\{ \left[B \cos(k_{\rm ABC} y) + 
    C \sin(k_{\rm ABC} z) \right] \hat{x} + \right. {} \nonumber \\
&& {} + \left[A \sin(k_{\rm ABC} x) + C \cos(k_{\rm ABC} z) \right] 
    \hat{y} + {} \nonumber \\
&& {} + \left. \left[A \cos(k_{\rm ABC} x) + B \sin(k_{\rm ABC} y) 
    \right] \hat{z} \right\},
\label{eq:ABC}
\end{eqnarray}}
\noindent with $k_{\rm ABC} = 3$, and $A=0.9$, $B=1$, and $C=1.1$. 
Previous results for $P_M<1$ have been reported in Refs. 
\cite{Brandenburg01,Archontis03}, although no systematic study of the 
effect of lowering $P_M$ was done.

The ABC flow is maximally helical, and the 
election of $k_{\rm ABC}$ allows for the growth of magnetic energy 
at scales larger than the energy injection scale. As for the Roberts flow, 
the ABC flow is also an exact solution of the Euler's equation, and 
turbulent fluctuations are generated as the result of a hydrodynamic 
instability at values of $R_V$ larger than a certain threshold. The 
particular election of the parameters $A$, $B$, and $C$ is done to 
break the symmetry of the flow and reduce the value of the 
hydrodynamic threshold \cite{Podvigina94,Archontis03}.

\subsection{Subgrid models of MHD turbulence}

As the value of $P_M$ is lowered down, the separation between the 
Ohmic and viscous dissipation scales increases. As a result, the velocity 
field has more small scale structure than the magnetic field. This 
imposes a stringent limit on the smallest $P_M$ that can be reached 
using DNS. At some point, the small scales in the flow cannot be 
solved anymore, and some form of SGS modeling is required.

While SGS models of hydrodynamic turbulence have a rich 
history, models for MHD flows are still in their infancy (see e.g. 
\cite{Muller02}). One of the main difficulties in MHD is that hypotheses 
often made in hydrodynamics (e.g. locality of interactions in Fourier space) 
are not necessarily true for magnetofluids. The general expression of the 
MHD energy spectrum is not known. And several regimes can be expected 
according to whether the system is mechanically or magnetically forced, 
whether the fields are statistically aligned or not, etc.

In this work we will focus on the LAMHD equations (or MHD 
$\alpha$-model). The hydrodynamic $\alpha$-model 
(see e.g \cite{Holm02a,Holm02b} and references therein) was validated 
against simulations in Ref. \cite{Chen99}. It differs from large eddy 
simulations (LES) in that the filter is applied to the Lagrangian of 
the ideal fluid, and as a result Hamiltonian 
properties of the system are preserved. It was extended to ideal MHD 
\cite{Holm02a,Holm02b}, and tested against DNS of MHD turbulence 
\cite{Mininni05a,Mininni05b}. In the context of magnetoconvection, the 
LAMHD equations were also studied in Ref. \cite{Jones05}. In the 
incompressible case, the LAMHD equations are
\begin{eqnarray}
\frac{\partial {\bf v}}{\partial t} + {\bf u}_s \cdot \nabla {\bf v} &=& 
    - v_j \nabla u_s^j -\nabla \widetilde{{\cal P}} + {\bf j} \times 
    {\bf B}_s \nonumber \\
    {} && + \nu \nabla^2 {\bf v} + {\bf f}, \label{eq:alpNS} \\
\frac{\partial {\bf B}_s}{\partial t} + {\bf u}_s \cdot \nabla {\bf B}_s 
    &=& {\bf B}_s \cdot \nabla {\bf u}_s + \eta \nabla^2 {\bf B} . 
    \label{eq:alpind} .
\end{eqnarray}
The pressure $\widetilde{{\cal P}}$ is to be determined, as before, 
from the relevant Poisson equation. The energy in this system is 
given by 
$E = \int ({\bf v} \cdot {\bf u}_s + {\bf B} \cdot {\bf B}_s)/2 \, d^3 x$. 
The subindex $s$ denotes smoothed fields, related 
to the unsmoothed fields by
\begin{eqnarray}
{\bf v} &=& \left(1 - \alpha^2 \nabla^2\right) {\bf u}_s \\
{\bf B} &=& \left(1 - \alpha^2 \nabla^2\right) {\bf B}_s . 
\end{eqnarray}

In the context of the dynamo at low $P_M$, some 
others SGS models have been used. Ref. \cite{Ponty05} used a modified 
LES \cite{Ponty04} where only the velocity field at scales smaller 
than the magnetic diffusion scale was modeled using a turbulent 
effective viscosity dependent on the wavenumber. A similar SGS 
model has been used in Ref. \cite{Schekochihin05} to study 
dynamo action with delta-correlated in time random forcing, but with 
the expression of the effective viscosity based on the 
Smagorinsky-Lilly model. Ref. \cite{Schekochihin05} 
also used hyperviscosity, although it should be remarked that this 
method is known to give wrong growth rates for the magnetic energy 
in the anisotropic case \cite{Grote00}.

\section{\label{sec:thresholds}BEHAVIOR OF THE DYNAMO WITH $P_M$}

\begin{figure}
\includegraphics[width=9cm]{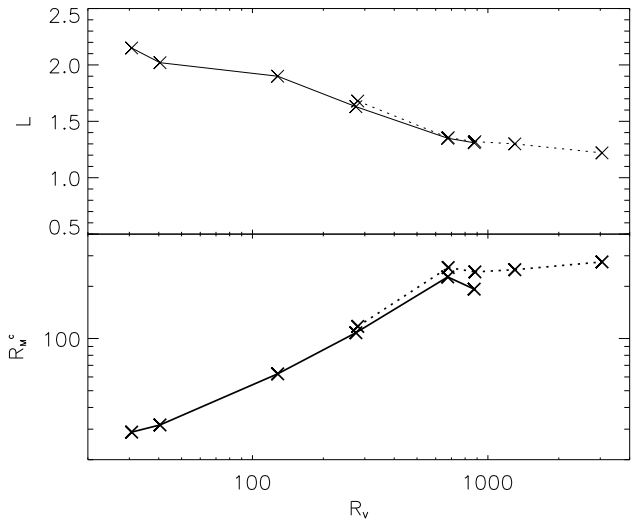}
\caption{Integral lengthscale of the flow as a function of $R_V$ 
     (above), and critical magnetic Reynolds for dynamo action $R_M^c$ 
     as a function of $R_V$ (below) for the Taylor-Green forcing. 
     Crosses connected with solid lines are obtained from DNS, while 
     crosses connected with dotted lines are obtained using the LAMHD 
     equations.}
\label{fig:TG}
\end{figure}

Figure \ref{fig:TG} (bottom panel) shows the critical magnetic Reynolds 
number $R_M^c$ as a function of $R_V$ for Taylor-Green forcing. Crosses 
connected with solid lines are obtained using DNS, while crosses 
connected with dotted lines are from LAMHD simulations (the largest 
resolution used by each method was of $512^3$ grid points). An overlap 
of the two methods for three values of $R_V$ was used to verify the 
SGS model. In Ref. \cite{Ponty05} higher values of $R_V$ were reached 
for this flow using LES.

Above the threshold, magnetic field excitations are amplified exponentially 
until the system reaches a fully developed MHD regime. Below the threshold, 
magnetic field perturbations are damped and the system reaches after long 
times a hydrodynamic regime. 

The threshold is obtained from simulations in the following way: in 
the steady state of a hydrodynamic simulation at a given value of $R_V$, 
a small magnetic field is introduced and several simulations varying 
$R_M$ are carried. In each MHD simulation, the exponential growth or 
decay rate $\sigma = d \log E_M /dt$ of the magnetic energy is 
measured. The value of $R_M$ for which $\sigma=0$ defines $R_M^c$. 
In practice, $R_M^c$ is bounded between two values of $R_M$ that 
give positive and negative $\sigma$, and the actual value of 
$R_M^c$ is obtained from a best fit to the $\sigma$ vs. $R_M$ curve 
(see e.g. Refs. \cite{Ponty05,Mininni05c}). This procedure gives the 
value of $R_M^c$ with an error of the order of $10\%$. 

Figure \ref{fig:TG} (top panel) also shows the flow integral scale 
as $R_V$ is increased. This scale is measured in the steady 
state of the hydrodynamic run that preceeds the MHD simulations. 
At low $R_V$ the flow is laminar and 
$L$ is close to the forcing scale $2\pi/k_{\rm TG}$. 
However, as $R_V$ is increased, more and more small scale excitations 
appear in the flow, and $L$ decreases. For $R_V \approx 1000$, the 
system has reached a fully 
turbulent regime with a Kolmogorov's energy spectrum. For larger values 
of $R_V$, $L$ changes slowly. Note that $R_M^c$ is 
anti-correlated with $L$. As $R_V$ increases, $R_M^c$ grows sharply. 
But for $R_V \approx 1000$ ($P_M \approx 0.3$) the threshold in $R_M^c$ 
settles around $300$, and the system seems to reach an asymptotic 
regime.

\begin{figure}
\includegraphics[width=9cm]{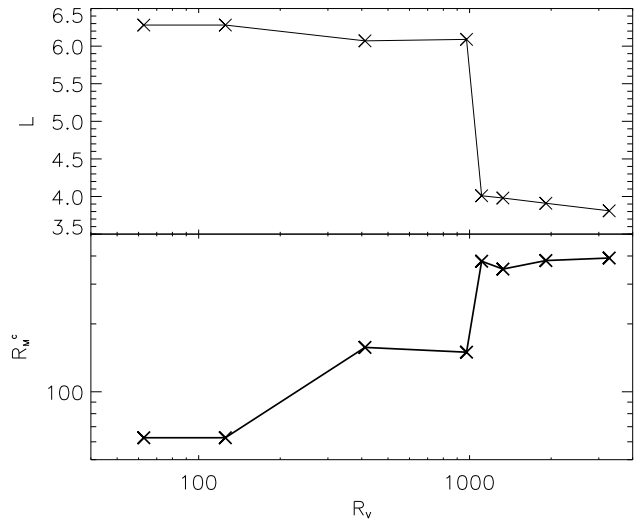}
\caption{Integral lengthscale of the flow as a function of $R_V$ 
     (above), and $R_M^c$ as a function of $R_V$ (below) for Roberts 
     forcing.}
\label{fig:GOR}
\end{figure}

The same quantities are shown in Fig. \ref{fig:GOR} for the Roberts forcing 
(see also Ref. \cite{Mininni05e}). In this case, the 
external forcing is an exact solution of the Euler's equation, and 
for $R_V \lesssim 100$ the flow is hydrodynamically stable. Between 
$R_V \approx 100$ and $R_V \approx 1000$ the flow displays 
hydrodynamic oscillations, and for $R_V \gtrsim 1000$ the flow reaches a 
hydrodynamic turbulent regime with a Kolmogorov's power law in 
the inertial range. As a result of these 
instabilities, the characteristic scale of the flow changes 
twice sharply. In the laminar regime $L \approx 2\pi/k_{\rm R}$. Then 
$L$ decreases and remains approximately constant in the oscillatory 
regime ($400\lesssim R_V \lesssim1000$), and finally has a second sharp 
decrease ($R_V \approx 1000$) 
as turbulence develops. The threshold for dynamo action shows again 
an anti-correlation with $L$, and two sharp increases in $R_M^c$ as a 
function of $R_V$ are observed as $L$ changes. In the turbulent 
regime ($R_V \approx 1000$), an asymptotic value of $R_M^c$ seems 
again to be reached. In this case, only DNS was used.

\begin{figure}
\includegraphics[width=9cm]{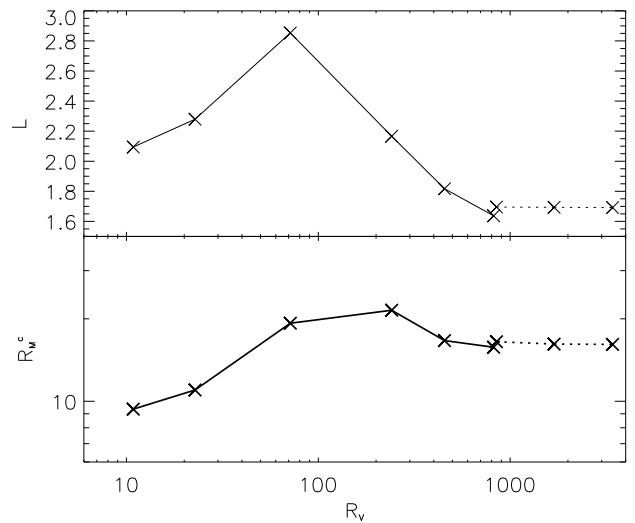}
\caption{Integral lengthscale of the flow as a function of $R_V$ 
     (above), and $R_M^c$ as a function of $R_V$ (below) for ABC 
     forcing. Crosses connected with solid lines are obtained from 
     DNS, while crosses connected with dotted lines are obtained 
     using the LAMHD equations.}
\label{fig:ABC}
\end{figure}

Figure \ref{fig:ABC} shows the threshold and integral scale 
as a function of $R_V$ for ABC forcing. As mentioned in the previous 
section, this is a helical flow with energy injection at intermediate 
scales, and as a result large scale dynamo action can take place.
As for Taylor-Green forcing, crosses connected with solid lines are 
obtained from DNS, while crosses connected with dotted lines are 
obtained from LAMHD simulations. There is an anti-correlation 
between $L$ and $R_V$, but it is less clear than in the previous flows. 
In the laminar case ($R_V \approx 10$) the integral scale is close to 
$2\pi/k_{ABC}$, while in the turbulent regime ($R_V \approx 1000$) 
$R_M^c$ is larger and $L$ is smaller. However, there is an intermediate 
range where $L$ grows together with $R_M^c$. The increase of $L$ in 
this intermediate regime is associated with the development of chaotic 
trajectories connecting several ABC cells as the flow bifurcates. The 
hydrodynamic bifurcations and subsequent generation of small scales 
(breaking the infinite time correlation of the laminar flow) also gives rise 
to an enhanced diffusivity that in turns increases $R_M^c$.

When compared with the 
previous flows, there are two striking differences in the amplitude 
of $R_M^c$ as a function of $R_V$. First, the thresholds at high $R_V$ 
are reduced by one order of magnitude. Second, there is only a small 
dependence of $R_M^c$ with $R_V$. As the flow destabilizes 
and develops turbulence ($R_V \approx 100$) the threshold changes 
from $R_M^c \approx 10$ to $R_M^c \approx 20$, while in the previously 
discussed flows the increase is by a factor of 10. After that, $R_M^c$ 
drops and finally stabilizes. Note that the smallest magnetic Prandtl 
number of $P_M \approx 5\times10^{-3}$ was reached for this flow.

\begin{figure}
\includegraphics[width=6cm]{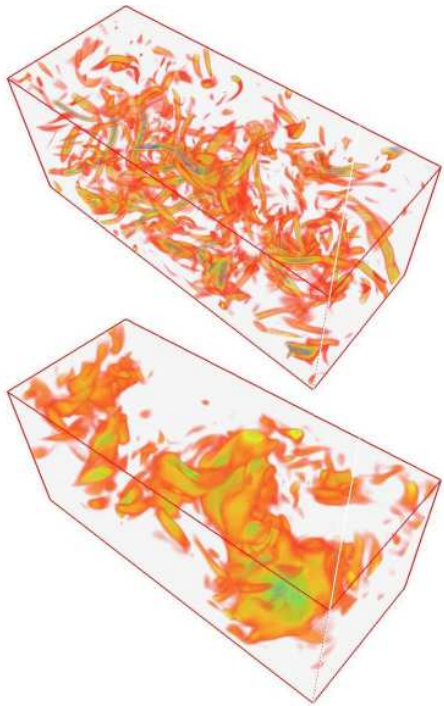}
\caption{Volume render of enstrophy density (above) and square 
     current density (below) in a small region of a simulation 
     with $P_M = 5\times10^{-2}$ and ABC forcing.}
\label{fig:render}
\end{figure}

Figure \ref{fig:render} shows renders of the enstrophy density and 
mean square current in a small box in a DNS using ABC forcing and 
$P_M = 5\times10^{-2}$. The scale separation between the kinetic 
and magnetic diffusion scales is evident. While the velocity field 
displays thin and elongated vortex tubes, the structures in the magnetic 
field are thicker.

In simulations without large scale flows 
\cite{Haugen04,Schekochihin05} no asymptotic regime for low 
$P_M$ has been found so far. As $R_V$ is increased and the 
system develops turbulent fluctuations, $R_M^c$ 
grows to values much larger than the ones obtained here. From the 
numerical results, it is not clear 
whether an asymptotic regime will be reached or $R_M^c$ will continue 
growing. However, theoretical arguments \cite{Boldyrev04} suggest that 
$R_M^c$ should be constant for values of $R_V$ large enough. 
From the comparison with the cases studied here and the anti-correlation 
found between $L$ and $R_M^c$, it is clear that 
the presence of a large scale flow (helical or not) plays a crucial 
role in the development of the asymptotic regime at relatively small 
values of $R_V$ and $R_M$.

\section{\label{sec:scales}SCALE INTERACTIONS}

The results presented in the previous section raise the question of 
what are the effects of turbulence, beyond increasing the value of 
$R_M^c$. Does dynamo action only take place at large scales 
when $P_M<1$, or can turbulent fluctuations also amplify small scale 
magnetic fields? By small scales, we refer here to scales smaller 
than the flow integral scale but larger than the magnetic diffusion scale.

\begin{figure}
\includegraphics[width=9cm]{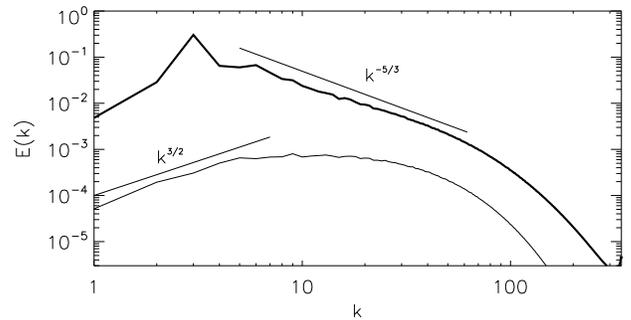}
\caption{Spectrum of kinetic energy (thick line) and magnetic 
     energy (thin line) in a $1024^3$ simulation with Taylor-Green 
     forcing ($P_M = 0.1$ and $R_M=400$). The spectrum of magnetic 
     energy has been multiplied by 10. Kolmogorov's and Kazanstsev's 
     power laws are shown as a reference.}
\label{fig:TGspectrum}
\end{figure}

Figure \ref{fig:TGspectrum} shows the kinetic and magnetic energy 
spectrum in the kinematic regime of a $1024^3$ DNS using 
Taylor-Green forcing ($P_M=0.1$ and $R_M=400$). 
The magnetic energy spectrum peaks at scales 
smaller than the flow integral scale, and at intermediate 
scales the spectrum is approximately flat. At large scales, a slope 
compatible with Kazantsev's spectrum \cite{Kazantsev68} is observed, 
although it should be noted that the hypothesis used by Kazantsev 
do not apply to this case. In Ref. \cite{Ponty05} it was shown that the 
peak in the magnetic energy spectrum moves to small scales as 
$R_V$ is increased.

The role played by the different scales in the flow is further clarified 
by examining the energy transfer functions. From Eqs. 
(\ref{eq:momentum}) and (\ref{eq:induc}), the transfers
\begin{equation}
T_L(k) = \int{\widehat{\mathbf v}_{\mathbf k}^* \cdot 
    \left(\widehat{{\mathbf j} \times {\mathbf B}} \right)_{\mathbf k} 
    d\Omega_{\mathbf k}} ,
\label{eq:TL}
\end{equation}
\begin{equation}
T_M(k) = \int{\widehat{\mathbf B}_{\mathbf k}^* \cdot \nabla \times 
    \left(\widehat{{\mathbf v} \times {\mathbf B}} \right)_{\mathbf k} 
    d\Omega_{\mathbf k}} ,
\label{eq:TM}
\end{equation}
can be defined (see e.g. Ref. \cite{Mininni05c}). The asterisk denotes 
complex conjugate, the hat Fourier transform, the subindex denotes the 
amplitude of the mode with wavevector ${\mathbf k}$, and 
$d\Omega_{\mathbf k}$ is the element of surface of the sphere of 
radius $k$ in Fourier space. When negative, $T_L(k)$ represents 
energy given by the velocity field at the 
wavenumber $k$ to the magnetic field at all scales. Positive $T_M(k)$ 
represents energy received by the magnetic field at wavenumber 
$k$  from the velocity field at all scales, and from the direct cascade of 
energy. The detailed shell-to-shell energy transfer in the kinematic 
dynamo has been studied in Ref. \cite{Mininni05d}.

\begin{figure}
\includegraphics[width=9cm]{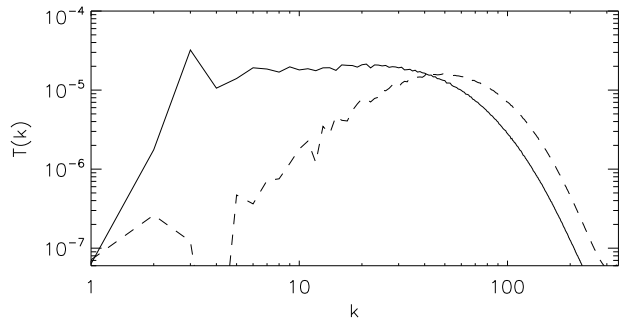}
\caption{Transfer functions $-T_L(k)$ (solid line) and $T_M(k)$ 
     (dashed line) in the $1024^3$ simulation with Taylor-Green 
     forcing.}
\label{fig:TGtransfer}
\end{figure}

Figure \ref{fig:TGtransfer} shows the transfer functions for the 
$1024^3$ DNS with Taylor-Green forcing. The magnetic field 
receives energy from the velocity field in all scales in the inertial 
range (note that $T_L$ has constant amplitude from $k \approx 3$ 
up to $k \approx 40$). As a result, for this forcing both the large scale 
flow and the turbulent fluctuations give energy to the magnetic field. 
On the other hand, $T_M(k)$ peaks at 
$k \approx 60$. This suggests that the dynamo process is non-local 
in Fourier space. Other indications of non-local interactions from 
simulations were obtained in Ref. \cite{Kida91}, and in simulations of 
dynamo action at large $P_M$ \cite{Schekochihin04}, although direct 
verification of the nonlocality in MHD turbulence was not obtained until 
recently \cite{Alexakis05,Debliquy05}.

Similar results were obtained for Roberts forcing \cite{Mininni05e}. In 
that case, a magnetic energy spectrum peaking at small scales and 
an approximately flat $T_L(k)$ transfer function at small scales were 
observed.

\begin{figure}
\includegraphics[width=9cm]{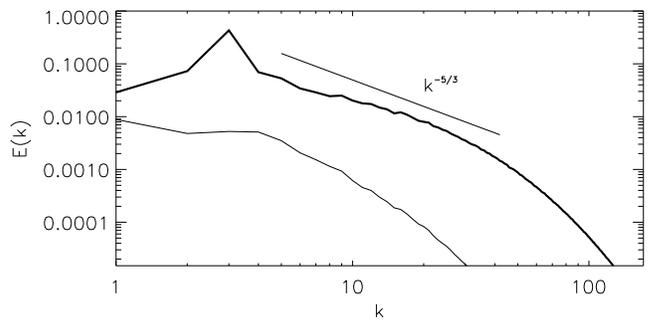}
\caption{Spectrum of kinetic energy (thick line) and magnetic 
     energy (thin line) in a $512^3$ simulation with ABC forcing 
     ($P_M = 1.5\times10^{-2}$ and $R_M=40$). The spectrum of 
     magnetic energy has been multiplied by 10.}
\label{fig:ABCspectrum}
\end{figure}

\begin{figure}
\includegraphics[width=9cm]{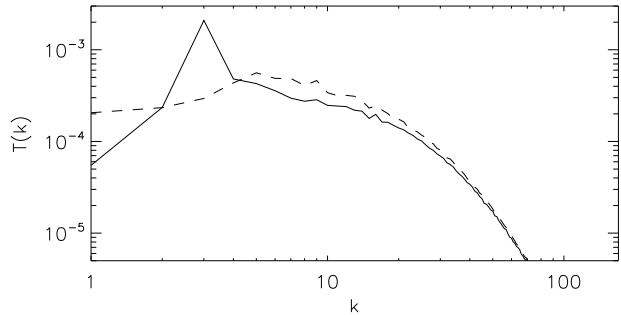}
\caption{Transfer functions $-T_L(k)$ (solid line) and $T_M(k)$ 
     (dashed line) in the $512^3$ simulation with ABC forcing.}
\label{fig:ABCtransfer}
\end{figure}

The results for ABC forcing are different. Figure \ref{fig:ABCspectrum} 
shows the kinetic and magnetic energy spectrum in a $512^3$ DNS 
with $P_M = 1.5\times10^{-2}$ and $R_M=40$. In this case, the 
magnetic energy spectrum peaks at $k=1$, and no peak is observed at 
scales smaller than the integral scale of the flow. Moreover, the magnetic 
energy spectrum drops fast. The transfer $T_L(k)$ (Fig. 
\ref{fig:ABCtransfer}) shows that most of the amplification is done by the 
large scale flow at $k=3$, and the turbulent fluctuations seem to give 
only a small contribution to the dynamo. $T_M(k)$ shows that most of the 
energy is received by the magnetic field at large scales, and it drops as 
$T_L(k)$ at small scales. 

The reason for this behavior can be understood as follows. For ABC 
forcing, $R_M^c$ is of the order of a few tens, 
even in the fully turbulent regime. As a result, a dynamo simulation 
just above threshold has small $R_M$, and small magnetic scales 
are damped so fast that no inertial range can develop 
in the magnetic energy spectrum. As a result, the magnetic field is 
only amplified by the large scale flow and grows at 
larger scales. Since there is no dynamo source at scales 
smaller than the flow integral scale, the small scales are only 
fed by the direct cascade of energy and the magnetic energy 
spectrum cannot peak at these scales.

However, if a simulation using ABC forcing is done at larger values of 
$R_M$, dynamo amplification by the small scales is recovered and 
similar features than for Taylor-Green forcing are obtained 
\cite{Brandenburg01,Mininni05d}. This situation should be the one 
that prevails in astrophysics.

\section{\label{sec:conclusions}CONCLUSIONS}

We reviewed results of dynamo action at low magnetic Prandtl number 
\cite{Ponty05,Mininni05c,Mininni05e} for several mechanical forcing functions, 
including helical and non-helical flows, as well as small and large scale 
dynamo action. For all cases where a large scale flow is present, a 
similar behavior is obtained in the threshold for dynamo action: as 
$R_V$ is increased (or $P_M$ decreased), the value of $R_M^c$ 
increases sharply as turbulence develops and finally reaches an 
asymptotic regime independent of the value of $P_M$. The large scale 
flow plays an important role in the establishment of the asymptotic behavior, 
as shown by the anti-correlation between the characteristic length of the 
flow and $R_M^c$. As turbulence develops, the laminar flow 
creates small scales through hydrodynamic instabilities, and the large scale 
laminar flow loses its infinite correlation time. Then, both the large and 
small scale velocity fields amplify the magnetic field, giving rise to the 
asymptotic regime.

New results from simulations with ABC forcing present a distinctive behavior. 
Having the flow maximum kinetic helicity and permitting large scale dynamo 
action, the critical magnetic Reynolds is smaller than for the other two 
flows by one order of magnitude. Also, only a twofold increase in $R_M^c$ 
is observed as turbulence develops (in contrast to a tenfold increase for 
the other flows). As a result, dynamo simulations close to the threshold do 
not show small scale amplification (down to $P_M = 5\times 10^{-3}$), and 
only the large scale flow is responsible for the large scale dynamo action. 
However, as the value of $R_M$ is increased, as is expected for astrophysical 
and geophysical flows, small scale amplification in the kinematic regime is 
recovered.

Although the conditions in our simulations are idealized, we believe the 
existence of an asymptotic regime for $P_M<1$ has profound implications 
for laboratory experiments and modeling of astrophysical and geophysical 
dynamos. In most of these cases, a large scale flow (such as differential 
rotation) and turbulent fluctuations are known to be present.

\begin{acknowledgments}
The author is grateful for valuable discussions with A. Alexakis, D.D. Holm, 
D.C. Montgomery, J.-F. Pinton, H. Politano, Y. Ponty, and A. Pouquet. Support 
from NSF grant CMG-0327888 is acknowledged. Computer time provided by 
NCAR, PSC, Darthmouth, and NERSC.
\end{acknowledgments}

\end{document}